\documentclass[11pt,twoside]{article}
\usepackage{asp2014}
\usepackage{graphicx}

\aspSuppressVolSlug
\resetcounters

\begin{document}

\title{Which Hydrogen Balmer Lines Are Most Reliable for Determining White Dwarf Atmospheric Parameters?}
\author{Ross~E.~Falcon$^{1,2}$, G.~A.~Rochau$^2$, J.~E.~Bailey$^2$, T.~A.~Gomez$^1$, M.~H.~Montgomery$^1$, D.~E.~Winget$^1$, and T.~Nagayama$^2$
\affil{$^1$Department of Astronomy and McDonald Observatory, University of Texas, Austin, TX 78712, USA}
\affil{$^2$Sandia National Laboratories, Albuquerque, NM 87185-1196, USA}}

\paperauthor{Ross~E.~Falcon}{refalco@sandia.gov}{}{University of Texas}{Department of Astronomy and McDonald Observatory}{Austin}{TX}{78712}{USA}
\paperauthor{G.~A.~Rochau}{garochau@sandia.gov}{}{Sandia National Laboratories}{}{Albuquerque}{NM}{87185-1196}{USA}
\paperauthor{J.~E.~Bailey}{jebaile@sandia.gov}{}{Sandia National Laboratories}{}{Albuquerque}{NM}{87185-1196}{USA}
\paperauthor{T.~A.~Gomez}{gomezt@astro.as.utexas.edu}{}{University of Texas}{Department of Astronomy and McDonald Observatory}{Austin}{TX}{78712}{USA}
\paperauthor{M.~H.~Montgomery}{mikemon@astro.as.utexas.edu}{}{University of Texas}{Department of Astronomy and McDonald Observatory}{Austin}{TX}{78712}{USA}
\paperauthor{D.~E.~Winget}{dew@astro.as.utexas.edu}{}{University of Texas}{Department of Astronomy and McDonald Observatory}{Austin}{TX}{78712}{USA}
\paperauthor{T.~Nagayama}{tnnagay@sandia.gov}{}{Sandia National Laboratories}{}{Albuquerque}{NM}{87185-1196}{USA}

\begin{abstract}
Our preliminary results from laboratory experiments studying white dwarf (WD) photospheres show a systematic difference between experimental plasma conditions inferred from measured H$\beta$ absorption line profiles versus those from H$\gamma$.  One hypothesis for this discrepancy is an inaccuracy in the relative theoretical line profiles of these two transitions.  This is intriguing because atmospheric parameters inferred from H Balmer lines in observed WD spectra show systematic trends such that inferred surface gravities decrease with increasing principal quantum number, $n$.  If conditions inferred from lower-$n$ Balmer lines are indeed more accurate, this suggests that spectroscopically determined DA WD masses may be greater than previously thought and in better agreement with the mean mass determined from gravitational redshifts.
\end{abstract}

\section{Introduction}

White dwarf (WD) atmospheric parameters (i.e., effective temperature, $T_{\rm eff}$, and surface gravity, log\,$g$) are broadly fundamental for astrophysics.  They are used to determine individual WD ages, which put strict observational constraints, independent of cosmological models, on the ages of stellar populations within our Galaxy \citep{Winget87}.  As another example, they are the starting point for understanding the progenitors of Type Ia supernovae, whose light curves are used as {\it standard candles} to measure extragalactic distances \citep[e.g.,][]{Colgate79}, which allow for the observation of our accelerating Universe \citep{Riess98}.

The most widely used technique to determine WD atmospheric parameters, known as the {\it spectroscopic method}, compares observed WD spectra with synthetic spectra generated using atmosphere models \citep[e.g.,][]{Koester79,Bergeron92b}.  Theoretical line profiles are a main ingredient to these models, and the incorporation of the latest calculations by \citet{Tremblay09} into the models used by the WD community results in systematic increases in inferred $T_{\rm eff}$ and log\,$g$.  The significance of this impact is one of the motivations for our experiments performed at the {\it Z} Pulsed Power Facility at Sandia National Laboratories \citep{Matzen05}.

\section{Creating White Dwarf Photospheres in the Laboratory}

\citet{Falcon10b} first introduce the laboratory work that creates plasmas at WD photospheric conditions ($T_{\rm e}\sim1\,$eV, $n_{\rm e}\sim10^{17}\,$cm$^{-3}$) for the purpose of investigating the theoretical line profiles used in WD atmosphere models.  We describe the setup and the acquisition of the time-resolved spectroscopic data in \citet{Falcon13}.  This also places our experimental platform in context by comparing it to other experiments that measure line profiles.  Distinguishing features of our platform include using a radiation-driven plasma source and observing our plasma in absorption in addition to in emission.

\subsection{What Is Accessible in the Laboratory?}

Gas cells enable us to create plasmas of various compositions.  Hydrogen has been our focus thus far since it addresses DA stars, the most abundant WD subclass.  Our scoping experiments prove the viability of helium \citep{Falcon13b} as well as carbon and oxygen, which can address DB and Hot DQ stars, respectively.

\begin{figure}[!h]
\begin{center}
  \includegraphics[width=0.97\columnwidth]{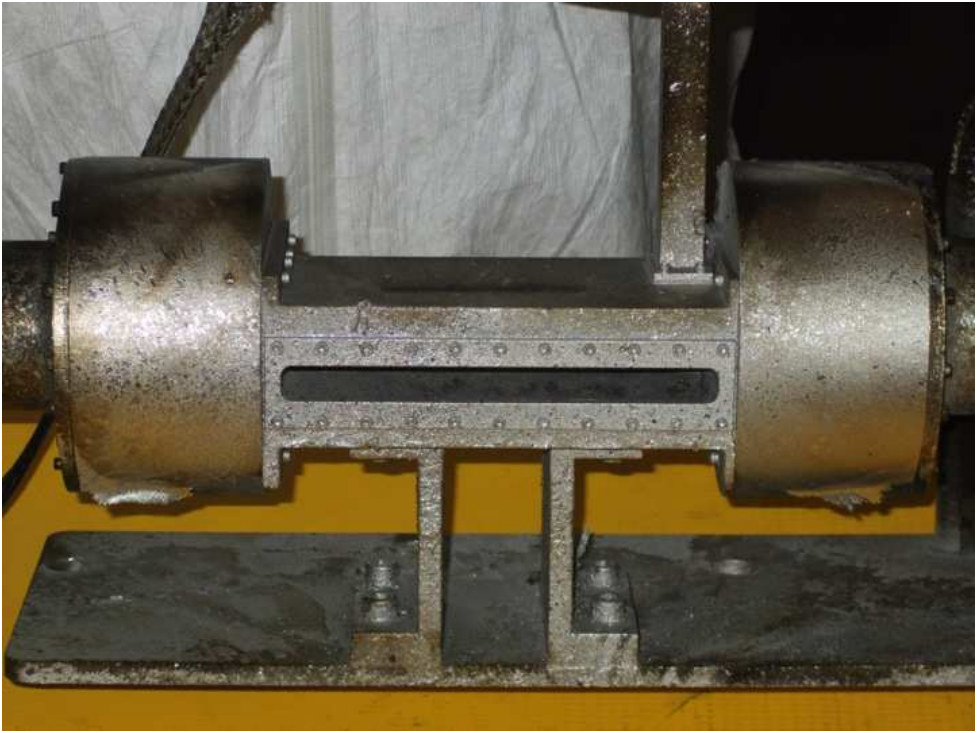}
  \caption{``ACE'' configuration gas cell that houses the hydrogen plasma during the experiment.  Capable of observing along three lines of sight, we simultaneously measure (1) our plasma in {\bf A}bsorption, (2) the {\bf C}ontinuum emission from a back-lighting surface, and (3) our plasma in {\bf E}mission.  In the aftermath of the pulsed power shot, debris and soot from the blast humble the appearance of the gas cell.}
  \label{before_fig}
\end{center}
\end{figure}

Our primary gas cell configuration uses three lines of sight to simultaneously observe: the plasma in {\bf A}bsorption, the back-lighting {\bf C}ontinuum, and the plasma in {\bf E}mission, lending to its name, ``ACE'' (Figure \ref{before_fig}).  Observing our laboratory plasmas in absorption affords us improved signal-to-noise (S/N) measurements compared to observing in emission only.  This is due to the high intensity of the back-lighter emission relative to the plasma self-emission and to the high population of the lower electronic energy level ($n=2$ for the Balmer series) relative to the upper level ($n=4$, 5, and 6 for H$\beta$, H$\gamma$, and H$\delta$, respectively).  This novel approach also constrains relative strengths among Balmer lines because they share the same lower level population, aiding not only our line shape measurements but potentially providing a technique to measure occupation probabilities \citep{Hummer88}.

Because our data are time-resolved, we monitor the evolution of our plasma.  We see its electron density, $n_{\rm e}$, smoothly increase with time as a consequence of its photoionized formation.  This means that we probe a range of $n_{\rm e}$ from the {\it same} plasma during a {\it single} experiment.

\subsection{The Preliminary Suggestion}

\begin{figure}[!h]
\begin{center}
  \includegraphics[width=\columnwidth]{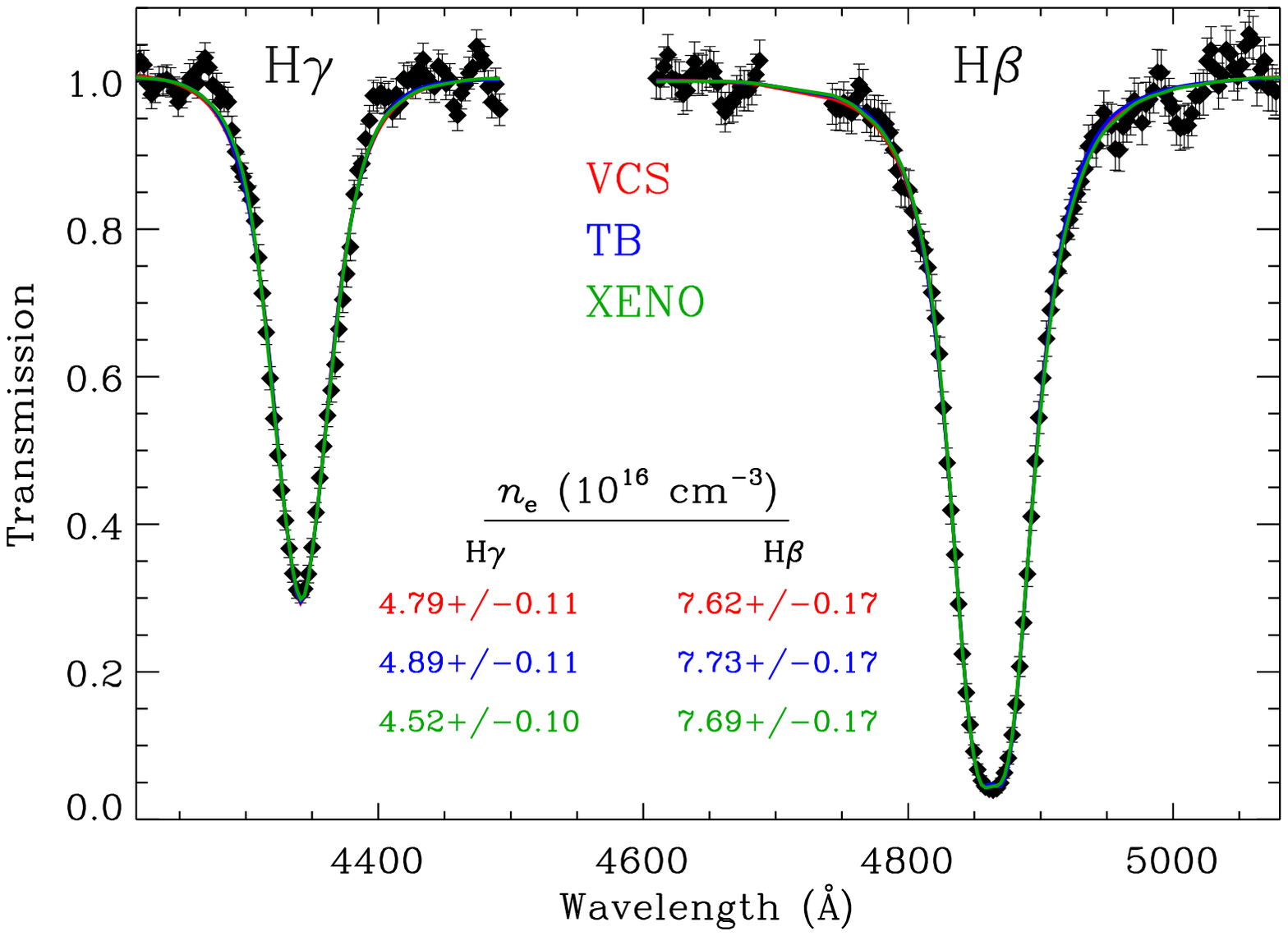}
  \caption{H$\beta$ and H$\gamma$ line transmission spectrum measured from our laboratory photoionized plasma and integrated over a 10-ns duration within our time-resolved data.  Red, blue, and green curves correspond to fits using line profiles following the theory of \citet[][VCS]{Vidal73}, \citet[][TB]{Tremblay09}, and \citet[][Xenomorph]{Gomez14}, respectively.  The fits overlap and return electron densities that are similar between theories but lower from H$\gamma$ than from H$\beta$ by $\sim40\,\%$.}
  \label{H4H5_fig}
\end{center}
\end{figure}

Figure \ref{H4H5_fig} plots an example of our measured H$\beta$ and H$\gamma$ line transmission (absorption divided by back-lighting continuum).  This spectrum is an integration over a 10-ns duration early on in the evolution of our time-resolved data, which covers $\sim300\,$ns.  Red, blue, and green curves correspond to fits using the theoretical line profiles of \citet{Lemke97} -- which follow the theory of \citet[][VCS]{Vidal73} -- \citet[][TB]{Tremblay09}, and \citet[][Xenomorph]{Gomez14}, respectively.  These fits assume $T_{\rm e}=1\,$eV, as motivated by simulations using the radiation-hydrodynamics code LASNEX \citep{Zimmerman78}.  From fitting H$\beta$ we infer $n_{\rm e}\sim7.7\times10^{16}\,$cm$^{-3}$ and from H$\gamma$ a value $\sim40\,\%$ less at $n_{\rm e}\sim4.7\times10^{16}\,$cm$^{-3}$.  We see a systematic underestimation of $n_{\rm e}$ inferred from H$\gamma$ versus H$\beta$ throughout the evolution of our experiment.

Without an independent plasma diagnostic, we cannot definitively say which set of inferred conditions are truly accurate, if either.  Even with another diagnostic, we likely cannot infer conditions to the precision necessary to discriminate between the H$\beta$ and H$\gamma$ line transmission fit determinations, since this line-fitting method is already so precise ($\sim4$ and 10\,\%  for H$\beta$ and H$\gamma$, respectively).

We have reasons to suspect that the H$\beta$ determination is more accurate than that from H$\gamma$.  H$\beta$ is a lower energy transition, so, theoretically, it is simpler to calculate, being less sensitive to high-$n_{\rm e}$ effects, such as continuum lowering \citep[e.g.,][]{Crowley13} and the mixing of electric field-dependent states \citep[e.g.,][]{Bethe57book}.  Experimentally, we observe H$\beta$ at larger optical depths.  This results in a higher S/N and more precise measurement than for H$\gamma$.  However, it also leaves open the possibility of saturation or large optical depth issues systematically compromising our measurement.  Preliminary investigations show no evidence that this is the case.  We continue to scrutinize our experiment to understand measurement uncertainties and to test our sensitivity to theoretical approximations in our analysis, such as assuming that $T_{\rm e}=1\,$eV.

A difference in experimental plasma conditions inferred from H$\beta$ versus from H$\gamma$ is interesting in the context of the spectroscopic method used in WD astronomy because of a long-standing problem in inferring different atmospheric parameters (i.e., $T_{\rm eff}$ and log\,$g$) from different H Balmer lines.  An ad hoc modification introduced by \citet{Bergeron93} to the occupation probabilities (line strengths) improves the consistency between conditions inferred from different Balmer lines when using VCS line profiles.  \citet{Tremblay09} further improve the consistency with their approach to calculating H line profiles and without using the ad hoc occupation probabilities.

However, all methods infer systematically lower surface gravities (log\,$g$) from lines with increasing principal quantum number, $n$.  This is the same as inferring lower $n_{\rm e}$ from higher-$n$ lines.

Fitting the H$\beta$ absorption line observed in WD spectra infers higher log\,$g$ and hence higher mass.  If the H$\beta$ determination is indeed more accurate, this results in a larger spectroscopic mean mass, which improves the agreement with the larger mean mass determined using the gravitational redshift method \citep{Falcon10}.

\section{A Call for a Different Look at WD Spectroscopic Data}

We call for an investigation of DA WD parameters spectroscopically inferred from different hydrogen Balmer lines and the comparison of those with parameters inferred independently from the spectroscopic method.  In particular, we are interested in parameters as a function of principal quantum number ($n$).  This will deteriorate the precision of the spectroscopic method, which usually relies on the inclusion of many lines.  Thus, this investigation may necessitate the analysis of ensemble characteristics or mean values for large samples of stars \citep[e.g.,][]{Falcon10}.

\acknowledgements  This work was performed at Sandia National Laboratories.  We thank the {\it Z} dynamic hohlraum, accelerator, diagnostics, materials processing, target fabrication, and wire array teams, without which we cannot run our experiments.  Sandia is a multiprogram laboratory operated by Sandia Corporation, a Lockheed Martin Company, for the United States Department of Energy under contract DE-AC04-94AL85000.  Thank you, P.-E. Tremblay, for providing the TB theoretical line profiles.  T.A.G acknowledges support from the National Science Foundation Graduate Research Fellowship under grant DGE-1110007.  M.H.M. and D.E.W. gratefully acknowledge support from the United States Department of Energy under grant DE-SC0010623.  This work has made use of NASA's Astrophysics Data System Bibliographic Services.

\bibliographystyle{asp2014}
\bibliography{/Users/falcon/Desktop/Astro/all}  


\end{document}